\documentclass[amsmath,amssymb, aps, pra,reprint,footinbib,floatfix,superscriptaddress]{revtex4-1}

\usepackage{float}
\usepackage{amsmath,amssymb}
\usepackage[english]{babel}
\usepackage{upgreek}
\usepackage{dsfont}
\usepackage{graphicx}
\usepackage{color}
\usepackage{bbold}
\usepackage{hyperref}
\usepackage[authormarkup=none]{changes}
\definechangesauthor[name={Stefan}, color=purple]{s}
\definechangesauthor[name={Bjoern}, color=blue]{b}
\definechangesauthor[name={Klaus}, color=red]{k}

\begin{document}
\title{Quantum-classical hypothesis tests in macroscopic matter-wave interferometry}

\author{Bj\"{o}rn Schrinski}
\affiliation{University of Duisburg-Essen, Faculty of Physics, Lotharstra\ss e 1, 47048 Duisburg, Germany}
\author{Stefan Nimmrichter}
\affiliation{Naturwissenschaftlich-Technische Fakult\"at, Universit\"at Siegen, Walter-Flex-Stra\ss{}e 3, 57068 Siegen, Germany}
\author{Klaus Hornberger}
\affiliation{University of Duisburg-Essen, Faculty of Physics, Lotharstra\ss e 1, 47048 Duisburg, Germany}

\begin{abstract}
We assess the most macroscopic matter-wave experiments to date
as to the extent to which they probe the quantum-classical boundary
by demonstrating interference of heavy molecules and cold atomic ensembles.
To this end, we consider a rigorous Bayesian test protocol for a parametrized set of hypothetical modifications of quantum theory, including well-studied spontaneous collapse models, that destroy superpositions and reinstate macrorealism. The range of modification parameters 
ruled out by the measurement events quantifies the macroscopicity of a quantum experiment, while the shape of the posterior distribution resulting from the Bayesian update reveals how conclusive the data are at testing macrorealism. This protocol may serve as a guide for the design of future matter-wave experiments ever closer to truly macroscopic scales.
\end{abstract}

\maketitle

\section{Introduction}

Matter-wave interference is one of the key observations that validate quantum mechanics and challenge macrorealism \cite{leggett1985quantum} and our classical perception of everyday life.
Attempts to explain the apparent absence of macroscopic superposition states, so-called Schr\"odinger cats, involve the many-world interpretation \cite{Everett1957}, decoherence theory \cite{joos2013decoherence,zurek2003decoherence,schlosshauer2007decoherence},
and
gravitational \cite{DIOSI1984199,Diosi1987} or spontaneous wavefunction collapse \cite{ghirardi1990markov,bassi2013models}. 
At the same time, longstanding experimental efforts 
are pushing the frontiers of quantum mechanics to ever larger 
spatial \cite{kovachy2015,asenbaum2017phase}, temporal  \cite{xu2019probing}, and mass  \cite{eibenberger2013,fein2019quantum} scales.
Systematic methods to assess and compare the prospects of 
testing macrorealism with different experimental approaches could guide future developments and allocation of resources in the field.

From a theory perspective, macroscopic quantum phenomena are often associated to 
quantum states whose macroscopic character manifests as a high degree of 
delocalization and 
entanglement in abstract many-body Hilbert space \cite{frowis2018}.
They are then 
gauged with functionals such as the quantum Fisher information \cite{Froewis2012}, that return large values for states intuitively deemed macroscopic.
This approach may lead to inconclusive or unintuitive results \cite{schrinskiassessing}. 
As an alternative, one may adhere to an empirical notion of macroscopicity
\cite{nimmrichter2014macroscopic}, which is based on how much the {observation} of quantum behavior constrains the hypothesis that quantum mechanics ceases to be valid on the macroscale. 

A common approach 
to quantum hypothesis testing 
is to 
consider the binary propositions that pure quantum {or} classical mechanics are more likely to have produced the observed measurement outcome \cite{Tsang2012,tsang2013testing,ralph2018dynamical}. In practice however, there are always unaccounted sources of noise and decoherence in the experiment so that both the quantum and the classical model 
are incomplete, and the measurement data will likely fit neither. 
One can alleviate this problem 
by instead considering a continuous hypothesis test against 
a set of 
\emph{minimal macrorealist modifications (MMM) } of quantum mechanics \cite{schrinski2019macroscopicity}. These models augment the Schr\"odinger equation by a 
parametrized stochastic process that destroys superpositions above a certain size, time, and mass threshold, while preserving them on the microscopic scale and fulfilling minimal consistency requirements \cite{Nimmrichter2013}. 
Macroscopicity then measures to what extent such classicalizing models are ruled out by the experimental demonstration of quantum effects.
The assessment of macroscopicity via a Bayesian hypothesis test refines the original formulation in Ref.~\cite{Nimmrichter2013} into an unambiguous definition.
Thanks to Bayesian consistency \cite{vaart1996weak,le2012asymptotics}, this agrees in the asymptotic limit of large amounts of data with consistent frequentist estimators based on expectation values. 

The formal definition of MMM and the continuous hypothesis test is summarized  in Sec.~\ref{sec:Empirical macroscopicity}. We then demonstrate how to employ this hypothesis test in the most relevant macroscopic matter-wave scenarios: near-field Talbot-Lau interferometry (Sec.~\ref{sec:Near field interferometry}) and atomic Mach-Zehnder interferometry (Sec.~\ref{sec:Mach-Zehnder interferometry}). After evaluating the 
macroscopicities of current record holders, we give a natural criterion in Sec.~\ref{sec:Hellinger} how to assess the amount of data needed to decisively rule out MMM, before concluding in Sec.~\ref{sec:Conclusion}.

\section{Empirical macroscopicity}\label{sec:Empirical macroscopicity}

A hypothetical modification of quantum mechanics that restores macrorealism while adhering to the fundamental symmetry principles of the theory should fulfill several consistency requirements \cite{Nimmrichter2013}. On a coarse-grained timescale all its observable consequences are captured by a Markovian extension of the von Neumann equation,
\begin{align}\label{eq:Superoperator}
\partial_t\rho=\frac{1}{i\hbar}[\mathsf{H},\rho]+\frac{1}{\tau_e}\mathcal{M}_\sigma\rho .
\end{align}
Here, the parameter $\tau_e$ sets the overall strength of the MMM effect, as given by the associated decoherence time for a single electron. The superoperator $\mathcal{M}_\sigma$ depends on a set $\sigma$ of additional parameters describing the details of the MMM. Complete positivity, Galilean covariance, and consistent many-body scaling requirements single out a particular Lindblad form of the MMM generator in second quantization \cite{Nimmrichter2013}, 
\begin{align}\label{eq:MMM}
\mathcal{M}_\sigma\rho=&\int d^3\mathbf{s}d^3\mathbf{q}\,
g_\sigma(s,q)\left[\mathsf{L}(\mathbf{q},\mathbf{s})\rho\mathsf{L}^\dagger(\mathbf{q},\mathbf{s})\right.\nonumber\\
&\left.-\frac{1}{2}\{\mathsf{L}^\dagger(\mathbf{q},\mathbf{s})\mathsf{L}(\mathbf{q},\mathbf{s}),\rho\}\right],
\end{align}
with
\begin{align}
\mathsf{L}(\mathbf{q},\mathbf{s})
=\sum_\alpha\frac{m_\alpha}{m_e}\int d^3\mathbf{p}\,
e^{i\mathbf{p}\cdot m_e\mathbf{s}/m_\alpha\hbar}
\mathsf{c}_\alpha^\dagger(\mathbf{p})\mathsf{c}_\alpha(\mathbf{p}-\mathbf{q}).
\end{align}
It applies to arbitrary many-body systems containing several  particle species $\alpha$ with masses $m_\alpha$. The electron   mass $m_e$ sets the reference scale. The $\mathsf{c}_\alpha (\mathbf{p})$ denote (fermionic or bosonic) particle annihilation operators in momentum representation. Applied to a single-particle state the   Lindblad operator $\mathsf{L}(\mathbf{q},\mathbf{s})$ effects a  phase-space translation on a characteristic scale set by the parameters $\sigma=(\sigma_q,\sigma_s)$, the momentum and position widths of the distribution $g_\sigma(s,q)$, which we assume to be Gaussian for simplicity. We avoid entering the relativistic regime by enforcing the upper bounds $\hbar/\sigma_q\geq 10\,\mathrm{fm}$ and $\sigma_s\leq20\,\mathrm{pm}$,  as discussed in Refs.~\cite{Nimmrichter2013,nimmrichter2014macroscopic}.
In practice, this renders the $\sigma_s$-scale irrelevant for all interferometer scenarios since they exhibit momentum superpositions far below $(m/m_e) \times \hbar / 20\,$pm.
The most prevalent instance of MMM by far, the continuous spontaneous localization (CSL) model \cite{ghirardi1990markov,bassi2013models}, is obtained by setting $\sigma_s=0$. 

MMM restore macrorealism by destroying coherences on a length scale determined by $\hbar/\sigma_q$, at an effective rate that amplifies with the mass and scales like $1/\tau_e$. Matter-wave interferometers or other mechanical superposition experiments then falsify MMM parameters $(\tau_e,\sigma)$ if the predicted coherence loss is incompatible with, say, the observed interference visibility. An experiment can thus be deemed as more macroscopic than another one if its measurement record falsifies a greater set of parameters. This empirical notion of macroscopicity can be cast into a quantitative measure using the concept of Bayesian hypothesis testing \cite{schrinski2019macroscopicity}. To this end, we consider the odds ratio
\begin{align}
o(\tau_e^*|d,\sigma,I)=\frac{P(H_{\tau_e^*}|d,\sigma,I)}{P(\overline{H}_{\tau_e^*}|d,\sigma,I)}
\end{align}
that quantifies wether a hypothesis $H_{\tau_e^*}$ or its rival hypothesis is more plausible, given the data $d$ accrued in an experiment and additional background information $I$ (which includes experimental parameters). 
Here, $H_{\tau_e^*}$ states that macrorealism holds and a MMM affects \eqref{eq:Superoperator} with a time parameter $\tau_e\leq\tau_e^*$ (at fixed $\sigma$); $\overline{H}_{\tau_e^*}$ assumes weaker modifications, $\tau_e > \tau_e^*$, or possibly none at all ($\tau_e \to \infty$). 
With help of Bayes' theorem, we can express the odds ratio in terms of the likelihoods $P(d|\tau_e,\sigma,I)$ that the MMM model \eqref{eq:Superoperator}  predicts the observed data $d$ based on the given parameters \cite{schrinski2019macroscopicity},
\begin{align}\label{eq:oddsratio1}
o(\tau_e^*|d,\sigma,I)=\frac{\int_0^{\tau_e^*} d\tau_e\,P(d|\tau_e,\sigma,I)p(\tau_e|\sigma,I)}{\int_{\tau_e^*}^\infty d\tau_e\,P(d|\tau_e,\sigma,I)p(\tau_e|\sigma,I)}.
\end{align}
We are left with specifying a prior probability $p(\tau_e|\sigma,I)$ for the MMM time parameter given the experimental scenario. A natural choice is Jeffreys' prior, defined as the square root of the Fisher information \cite{jeffreys1998theory} of the likelihood with respect to $\tau_e$,
\begin{align}\label{eq:JeffreysPrior}
p(\tau_e|\sigma,I)&\sim\sqrt{\mathcal{F}(\tau_e|\sigma,I)}\nonumber\\
&=\sqrt{\left\langle\left(\frac{\partial}{\partial\tau_e}\mathrm{log}[P(D|\tau_e,\sigma,I)]\right)^2\right\rangle_D},
\end{align}
 where $\left\langle\cdot\right\rangle_D$ is the expectation value with respect to the sample space $D$ of elementary experimental outcomes, from which the $d$ are drawn. The prior \eqref{eq:JeffreysPrior} is the most objective choice when it comes to comparing different experiments. Jeffreys' prior is invariant under reparametrizations and results in the largest information gain between prior and posterior on average, as measured by the Kullback-Leibler divergence (or relative entropy) \cite{schrinskiassessing,bernardo1979reference,ghosh2011objective}.

As a conservative criterion for rejecting the MMM hypothesis $H_{\tau_e^*}$ we require the odds ratio \eqref{eq:oddsratio1} to fall below 1:19. This is equivalent to determining the lowest five-percent quantile of the posterior distribution $P(d|\tau_e,\sigma,I)p(\tau_e|\sigma,I)/P(d|\sigma,I)$. The respective time parameter $\tau_\mathrm{m}$ marking this quantile then defines the empirical macroscopicity,
\begin{align}\label{eq:mu}
\mu_\mathrm{m}=\max_\sigma\left[\mathrm{log}_{10}\left(\frac{\tau_\mathrm{m}(\sigma)}{1\,\mathrm{s}}\right)\right].
\end{align}
It characterizes the MMM time parameters most probably ruled out by the data. By  maximizing over all $\sigma$ one obtains a single figure of merit.

\section{Near-field interferometry}\label{sec:Near field interferometry}

Molecule interferometers are 
prone to achieve  high macroscopicities  due to the large masses involved. Recently, the long-baseline universal matter-wave interferometer (LUMI) \cite{fein2019quantum},
an extended version of the 
Kapitza-Dirac-Talbot-Lau  interferometer  (KDTLI) \cite{eibenberger2013},
demonstrated interference of individual molecules of more than $2.5\times 10^4$ atomic mass units. It makes use of the Talbot-Lau  near-field interference effect based on diffraction off a standing light wave.

The experimental apparatus consists of three equidistant gratings of equal grating period $d_\mathrm{g}$. The distance $L$ between them is of the order of the Talbot length, which marks the near-field regime \cite{Hornberger2009,hornberger2012colloquium}. An initially incoherent molecule beam is collimated by the first grating and then diffracted by the second one, which results in a periodic interference fringe pattern at the position of the third grating. In the KDTLI and the LUMI setups, the second grating consists of a standing laser wave that modulates the phase of the molecular wave,
while the first and third grating are material masks.
The interference pattern is scanned by varying the lateral position $x_{\rm S}$ of the third grating mask and counting the number of transmitted molecules as a function of $x_{\rm S}$. One obtains an approximately sinusoidal detection signal \cite{Hornberger2009,nimmrichter2014macroscopic}, 
\begin{align}\label{eq:MIintensity}
S(\tau_e,\sigma,I)=f_1 f_3 \left\{1+\mathcal{V}\,\mathrm{sin}\left[\frac{2\pi}{d_\mathrm{g}}(x_\mathrm{S}+\delta x)\right]\right\}.
\end{align}
Here, 
$f_{1},f_3$ denote the opening fractions of the first and third grating, 
and  $\delta x$ is the grating position offset, which is not measured but can be extracted from a fit to the data.

Quantum mechanics predicts an ideal interference contrast (or visibility) $\mathcal{V}_0$ that depends on various parameters such as the molecular mass $m$, the time of flight $T$, and the grating laser power $P_{\rm L}$ \cite{Hornberger2009} which are all part of the background information $I=\{x_\mathrm{S},P_\mathrm{L},\dots\}$.
MMM-induced decoherence will effect a reduction of the contrast according to \cite{Nimmrichter2011}
\begin{align} \label{eq:vis}
\frac{\mathcal{V}}{\mathcal{V}_0}=\exp\left\{-\frac{2Tm^2}{\tau_e m_e^2}\left[1-\frac{\sqrt{\pi}\hbar T_\mathrm{T}}{\sqrt{2}d_\mathrm{g}\sigma_qT}\mathrm{erf}\left(\frac{d_\mathrm{g}\sigma_qT}{\sqrt{2}\hbar T_\mathrm{T}}\right)\right]\right\}.
\end{align}
with $T_\mathrm{T}=m d_g^2/h$ the Talbot time. In practice, the velocity at which individual molecules are ejected from the source and traverse the setup is not known, 
so that the signal \eqref{eq:MIintensity} must be averaged over a measured time-of-flight distribution.

Previous  estimates of the achieved macroscopicity \cite{Nimmrichter2013,fein2019quantum}  compared the measured contrast with \eqref{eq:vis}, 
attributing a certain confidence to the latter, 
and deduced the greatest excluded $\tau_e$-value from there. Here we carry out a proper statistical analysis based on the raw molecule count data, which fully accounts for all measurement uncertainties and allows dealing with small noisy data sets. We distinguish two modes of measurement: stationary operation with a constant molecule flux (KDTLI), which requires 
an additional uniformity assumption for the count statistics, and pulsed operation (LUMI).

The probability of particles to end up at the detector is directly proportional to the intensity $S$ in \eqref{eq:MIintensity}.
Given the experimental setup, we may 
assume a constant particle flux that illuminates  homogeneously many grating slits and
an efficient detector that covers all these slits, i.e.~every molecule that passes the first grating and is not blocked by the third one will be counted. 
The probability for a molecule that has entered the interferometer to pass through all of its openings is then
\begin{align}\label{eq:MIlikelihood}
P(+|\tau_e,\sigma,I)=S(\tau_e,\sigma,I)/f_1,
\end{align}
given a fixed lateral position $x_{\rm S}$ of the third grating.

In the KDTLI experiment \cite{eibenberger2013}, the measurement record consists of the count numbers $N^+_{x_\mathrm{S}}$ of detected molecules at each $x_{\rm S}$. The 
complementary events with probability $P(-|\tau_e,\sigma,I) = 1-P(+|\tau_e,\sigma,I)$ are missing as blocked molecules cannot be detected. 
However, the numbers $N^-_{x_\mathrm{S}}$ of blocked molecules can be deduced from the sum of all counts, $N_\mathrm{tot} = \sum_{x_{\rm S}} N^+_{x_{\rm S}}$, with help of the \emph{fair-sampling hypothesis}, i.e.\,the position of the third grating does not influence the lateral probability density of the molecules.
To this end, we  use that the third grating uniformly scans the lateral dimension in $M$ equidistant steps extending over a multiple of the period $d_{\rm g}$. At constant flux, this implies that about the same number of molecules, $N_{x_\mathrm{S}}=N_\mathrm{tot} /M f_3$ 
must have traversed the interferometer in each step, out of which $N^-_{x_\mathrm{S}}=N_{x_\mathrm{S}}-N^+_{x_\mathrm{S}}$ were blocked.
The posterior for the MMM time parameter then reads as
\begin{align}
&p(\tau_e|d,\sigma,I)\propto  p(\tau_e|\sigma,I)\nonumber\\
&\times\prod_{x_\mathrm{S},P_{\rm L}}[S(\tau_e,\sigma,I)]^{N^+_{x_{\rm S},P_\mathrm{L}}}[f_1-S(\tau_e,\sigma,I)]^{N^-_{x_{\rm S},P_\mathrm{L}}},
\end{align}
omitting normalization.  
This expression also accounts for data taken at different values $P_{\rm L}$ of the grating laser power, as was done in \cite{eibenberger2013}. Further data could be simply appended to the 
product
if other  parameters are varied in consecutive grating scans, e.g. grating separations, time-of-flight distributions, or the molecular species. 

\begin{figure}
  \centering
  \includegraphics[width=0.45\textwidth]{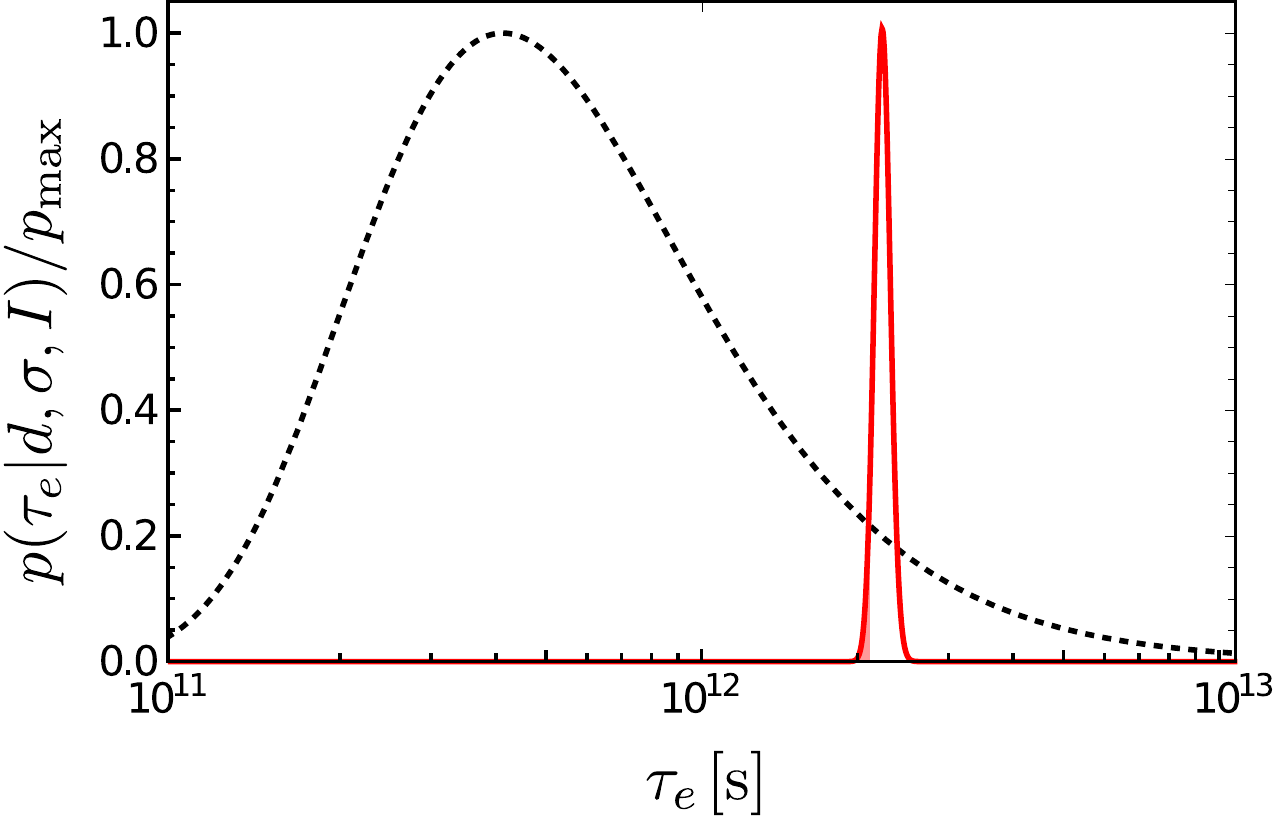}
  \caption{Analysis of the KDTLI experiment from Ref.~\cite{eibenberger2013}: Jeffreys' prior is shown as a dotted line for two combined experimental runs at
0.84\,W and 1\,W, the posterior after updating with the data is depicted by a solid line. All distributions are normalized to their maximum value $p_\mathrm{max}$ and
the lowest five percent quantile of the posterior is marked by a shaded area.}
\label{fig:KDTLI}
\end{figure}

Figure \ref{fig:KDTLI} shows the posterior after plugging in the data from Ref.~\cite{eibenberger2013}, using two runs at grating laser power $P_{\rm L}=0.84\,$W and $P_{\rm L}=1\,$W. 
For this we take the characteristic MMM length scale $\hbar/\sigma_q$ to  exceed by far the molecule size $d_\mathrm{m}$ and to stay  below the typical interference path separation, $d_\mathrm{m} \ll \hbar/\sigma_q \ll d_g T/T_\mathrm{T}$. 
This admits the point-particle description above while
maximizing the decoherence effect to $\mathcal{V}\simeq\mathcal{V}_0 \exp[-2 T m^2/\tau_e m_e^2]$.  For future experiments where the spatial dimension of the molecules might be of the order of the grating period a treatment beyond the point particle approximation will be necessary \cite{belenchia2019talbot}.

One observes that  Jeffreys' prior (dotted line) is updated to larger MMM time scales (i.e.~weaker modifications) and to a much narrower distribution. As  further discussed in Sec.~\ref{sec:Hellinger}, this already indicates a conclusive 
measurement 
and a good posterior convergence for these runs.  
The plot also illustrates the double
role of Jeffreys' prior in the assessment of macroscopicity: On the one hand, the prior gives a rough forecast of what MMM time scales one can access and what macroscopicity one can expect from a certain experimental setup with a given mass, time, and length scale. On the other hand, the overall performance of the experiment in terms of data quantity and quality decides whether the Bayesian update will ultimately converge to a sharp posterior distribution that no longer resembles the prior. We can then speak of a conclusive observation of genuinely macroscopic quantum behavior corresponding to the macroscopicity value $\mu_\mathrm{m}$. 

The LUMI experiment \cite{fein2019quantum} is an extended version of the KDTLI setup  that operates in a pulsed regime and is designed for more massive particles. During each shot, the laser grating is switched on and off;  the  molecules are detected in a time-resolved manner, resulting in two count numbers per shot, $N^+_{x_\mathrm{S}}$ and $N^0_{x_\mathrm{S}}$ with the laser on and off, respectively. The second value allows us to infer the number of blocked molecules required for the Bayesian update, $N^-_{x_\mathrm{S}} = N^0_{x_\mathrm{S}}/f_3 - N^+_{x_\mathrm{S}}$, since the probability for a molecule to pass the third grating is simply $f_3$ in the absence of a second grating \footnote{This assumes a sufficiently homogeneous 
particle pulse with no significant density fluctuations between the time windows of activated and deactivated laser, 
as realized in the experiment.}. The lateral position $x_{\rm S}$ is again varied from pulse to pulse, and the procedure is repeated for varying grating laser powers.

\begin{figure}
  \centering
  \includegraphics[width=0.45\textwidth]{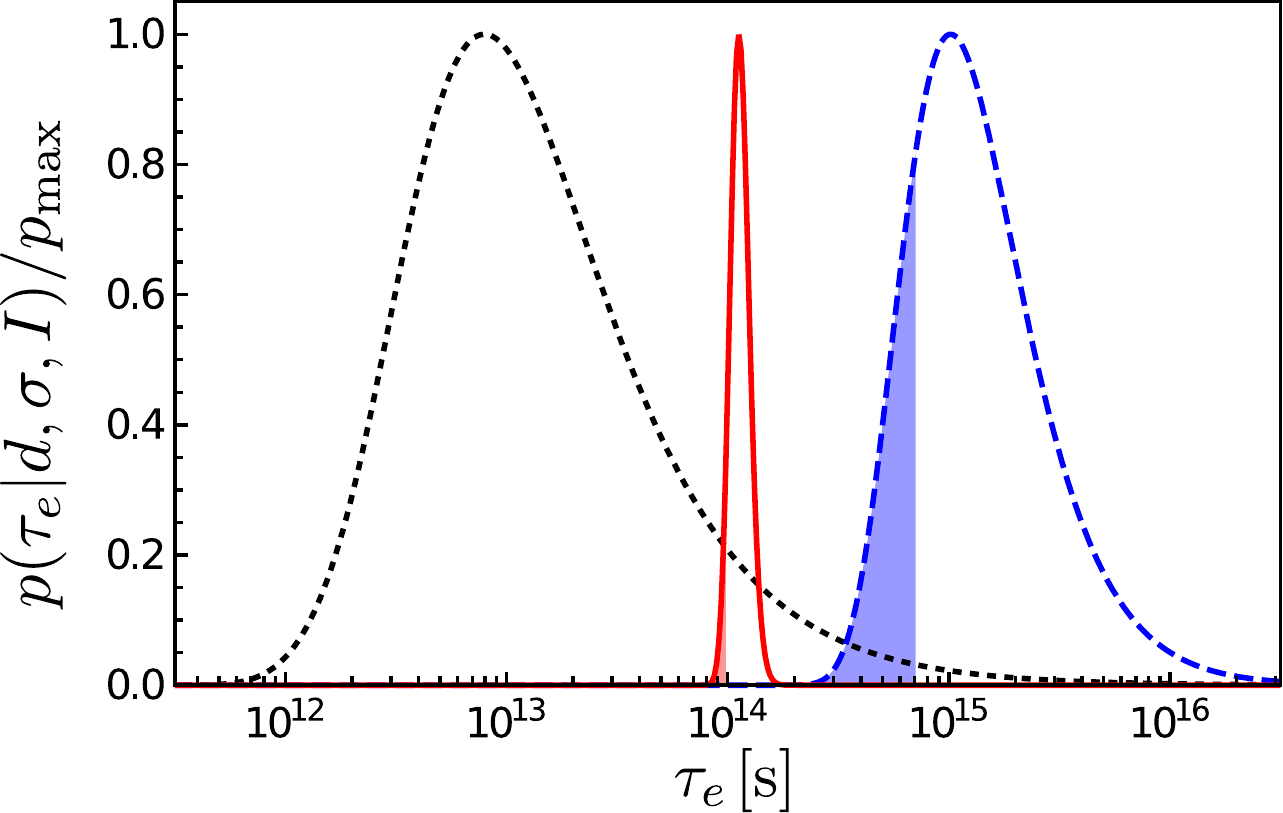}
  \caption{Analysis of the LUMI experiment from Ref.~\cite{fein2019quantum}: Shown are Jeffreys' prior (dotted line) together with the posterior distributions achieved by updating
the prior with results of the LUMI experiment \cite{fein2019quantum}: Using the eight best data sets leads to the blue posterior (dashed line) while using all data (grating laser power ranging from $P_{\rm L}=0.2$ to $1.8$\,W) leads to the red posterior (solid line). Jeffreys' prior coincides for both scenarios.  All distributions are normalized to their maximum value $p_\mathrm{max}$ and
the lowest five percent quantiles of the posteriors are marked by the shaded areas.}
\label{fig:LUMI}
\end{figure}

The measurement data from 23 LUMI runs at laser powers between $0.2$\,W and $1.8$\,W, taken from Ref.~\cite{fein2019quantum}, result in the posterior shown in Fig.~\ref{fig:LUMI} (red solid line). Similar to the KDTLI case, we obtain a sharply peaked, approximately Gaussian distribution around 
greater 
classicalization  
time scales, i.e. weaker MMM, which corresponds to the macroscopicity
$\mu_\mathrm{m}=14.0$. 

The individual LUMI runs at different laser powers comprise roughly an order of magnitude fewer molecule counts than in the KDTLI experiment. Thus, more of these runs have to be combined to reach the same level of posterior convergence. 
Nevertheless, one might be tempted to postselect
among
the 23 runs, discarding those with a poor interference visibility in order to achieve maximum macroscopicity. 
Indeed, by taking only the best eight runs into account, we can boost the macroscopicity to $\mu_\mathrm{m}=14.8$. But we are left with a broad posterior (blue dashed line in Fig.~\ref{fig:LUMI}) that has not yet converged towards a Gaussian shape and still resembles the prior (dotted line). Such an outcome suggests that the hypothesis test is based on too little data to be fully conclusive. In Section \ref{sec:Hellinger}, we will introduce a quantitative criterion for how conclusive a set of data is in terms of empirical macroscopicity.  

\section{Mach-Zehnder interferometry}\label{sec:Mach-Zehnder interferometry}

\begin{figure*}
  \centering
  \includegraphics[width=0.80\textwidth]{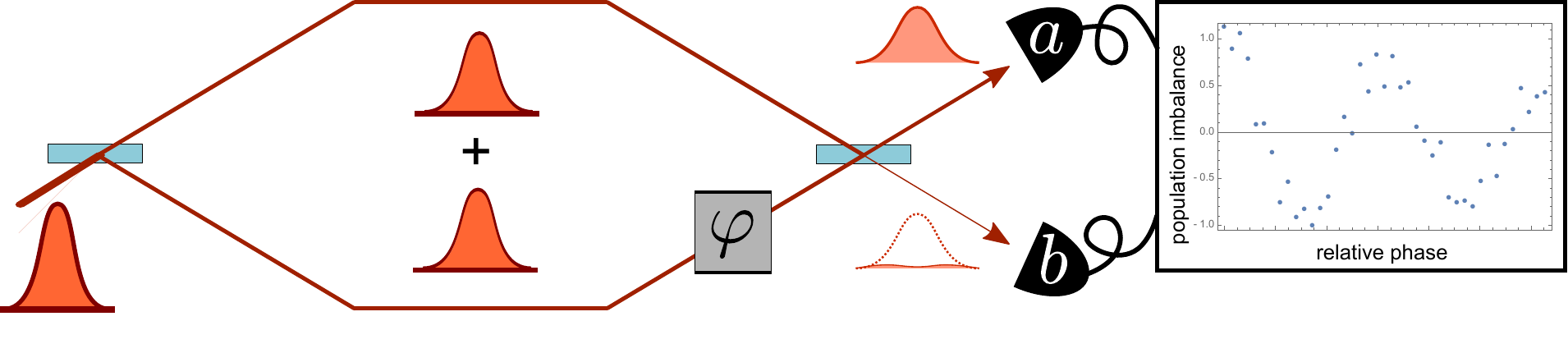}
  \caption{
    Typical setup of a Mach-Zehnder matter-wave interferometer: A first beam splitter prepares individual or Bose-Einstein condensed atoms in a momentum superposition,
    usually by means of an internal $\pi/2$-transition \cite{kovachy2015,xu2019probing}. A momentum inversion, or the implementation of an atom guide, makes them impinge on a second beam splitter. Interference is demonstrated by counting the atom number in the two output ports in dependence on the relative phase $\varphi$ between the modes, which may be tuned e.g. by varying the times of flight or a potential gradient. }
\label{fig:MachZehnder}
\end{figure*}

We now turn to interferometers of the Mach-Zehnder type, see Fig.~\ref{fig:MachZehnder}. A first beam splitter creates a superposition of two wave packets, occupied by a single atom or an entire BEC, which propagate on  distinct paths,  rejoin on a second beam splitter, and are then detected in the  two associated output ports.
Experimental demonstrations of such phase-stable superposition states 
may be deemed macroscopic due to the large arm separations and long coherence times that can be achieved
\cite{Schmiedmayer2013,kovachy2015,asenbaum2017phase,xu2019probing}.

One may distinguish two basic types of operation: Either individual, i.e.\ distinguishable, atoms are sent through the interferometer, or a BEC of many identical particles passes the setup,
whose 
macroscopic wave function then interferes with itself.
In the ideal case, where  no particles are lost, coherence and a stable phase
$\varphi$ are maintained, 
interactions can be neglected,
and the particles are detected with unit efficiency, both scenarios lead to the same binomial distribution for the number of atoms $n_a$ and $n_b=N-n_a$ recorded in output ports $a$ and $b$,
\begin{align} \label{eq:binomialCount}
p(n_a|N,\varphi) = {N \choose n_a} \cos^{2n_a}\left(\frac{\varphi}{2}\right)\sin^{2(N-n_a)}\left(\frac{\varphi}{2}\right).
\end{align}
This indicates that, in terms of empirical macroscopicity, BEC interference with thousands of atoms in a product state is equivalent to single-atom interference with the same number of repetitions. It turns out that even partial entanglement through squeezing \cite{sorensen2001many,Sorensen2001,Schmiedmayer2013} has no noticeable advantage for BECs 
regarding macroscopicity, as long as only collective observables are measured \cite{schrinski2019macroscopicity}. However, in the presence of decoherence effects and experimental disturbances
the equivalence of both scenarios no longer holds.

\subsection{Two-mode interference of BECs}\label{sec:tmbec}

Suppose a BEC is split coherently at the first beam splitter so that all atoms  share the same collective phase at the beginning. Even in this case, the interference pattern may fluctuate randomly from shot to shot due to 
technical noise in the beam splitters, timing uncertainties, or fluctuating background fields.
In case of full dephasing one expects the uniform phase average of the binomial count distribution \eqref{eq:binomialCount} 
as given by
\begin{align}\label{eq:discreteBECdecohered}
p(n_a|N) = & \frac{\Gamma(n_a+1/2)\Gamma(N-n_a+1/2)}{\pi\Gamma(n_a+1)\Gamma(N-n_a+1)}.
\end{align}
It is smeared out over the whole range of
$0\leq n_a \leq N$, with higher probabilities at the margins; the general expression for all 
stages of dephasing is reported in \cite{schrinski2017sensing}.

Notice the striking difference between \eqref{eq:discreteBECdecohered} and the binomial distribution peaked at $n_a=N/2$, which is predicted by a classical coin-flip model of the interferometer
where the beam splitters send individual atoms
into one or the other arm at random (see a specific example in App.\ \ref{AppA}). 
Recording atom count numbers $n_a$ far from $N/2$ in individual runs of the BEC interferometer therefore demonstrates a non-classical effect; however, since practically the same count statistics \eqref{eq:discreteBECdecohered} are expected if two separately prepared BECs 
are  sent simultaneously onto a beam splitter \cite{Laloe2012,schrinski2017sensing}
it is doubtful whether a genuine quantum superposition  can be confirmed at all without a stable phase \cite{stamper2016verifying}. 

The measure of macroscopicity introduced in this paper is particularly suited to  clarify this issue, as it based on a Bayesian hypothesis test of MMM that prevent superposition states.
Given that typically hundreds or thousands of
atoms arrive at the detectors, one can perform a continuum approximation to obtain a closed expression for the partially dephased atom count distribution \cite{schrinski2019macroscopicity},
 \begin{align}\label{eq:partdec}
p(n_a|\tau_e,\sigma,I)=&\frac{\Theta[\cos^2(\delta_a)]}{2\pi|\cos(\delta_a)|}\left[\vartheta_3\left(\frac{\delta_a-\varphi}{2},g_N(t)\right)\right.\nonumber\\
&+\left.\vartheta_3\left(\frac{\pi-\delta_a-\varphi}{2},g_N(t)\right)\right],
\end{align}
with $\delta_a\equiv\arcsin(2n_a/N-1)$. Dephasing with rate $\Gamma_\mathrm{P}$ is here captured by the function $g_N(t)=\exp\left[-1/2N-\Gamma_\mathrm{P}t/2\right]$ and 
$\vartheta_3$ is the Jacobi-theta function of the third kind, 
\begin{align}
\vartheta_3(u,q)=\sum_{n=-\infty}^\infty q^{n^2}e^{2inu}.
\end{align}
The distribution \eqref{eq:partdec} is an oscillatory function of the phase difference $\varphi$ between the two Mach-Zehnder arms. 
Interference visibility is reduced, in parts, by the initial phase uncertainty of the $N$-atom product state and by gradual dephasing over the interference time $t$.
MMM predict the dephasing rate
\begin{align}\label{eq:DephRate}
\Gamma_\mathrm{P}=\frac{2m^2}{\tau_e m_e^2}
\frac{1-\exp[-\Delta_x^2\sigma_q^2/(4\sigma_q^2 w_x^2+2\hbar^2)]}{\sqrt{(1+2\sigma_q^2 w_x^2/\hbar^2)(1+2\sigma_q^2 w_y^2/\hbar^2)}},
\end{align}
with $\Delta_x=\Delta_p T/2\sqrt{3}m$ the effective \emph{average} arm separation, given the momentum splitting $\Delta_p$ and the atom mass $m$. We assume free evolution of the BEC in the $z$-direction, a Gaussian transverse mode profile with waists $w_x,w_y$, and negligible mode dispersion; see \cite{schrinski2017sensing} for a detailed derivation. Given a highly diluted BEC, one may also neglect phase dispersion caused by atom-atom interactions \cite{Javanainen1997PhaseDispersion}. Its presence would further reduce the interference visibility, which would be falsely attributed to MMMs, and omitting it
thus underestimates macroscopicity.

Atoms lost from the condensate can simply be traced out \cite{Ma2011}, 
given that we have a product state of single-atom superpositions. Our reasoning thus applies to the remaining particle number $N$ registered in the detectors. In fact, MMM also lead to atom loss, and for $\sigma_q > \hbar/w_{x,y}$ this depletion dominates, 
while the dephasing rate \eqref{eq:DephRate} drops. The reason is simple: undetected atoms have no effect on the interference signal
\footnote{A fraction of the depleted atoms might still arrive at the detectors, which could be accounted for by adding a flat background to the signal.}. In principle, one could then rule out macrorealistic modifications such as the CSL model by testing their predicted depletion rates against actually measured atom loss over time \cite{Laloe2014}. But since no quantum signatures would be verified in such a scheme, no macroscopicity should be assigned to such an observation either.   
As a genuine quantum experiment, the BEC interferometer is most sensitive to MMM dephasing in the parameter range $w_{x,y} \ll \hbar/\sigma_q \ll \Delta_x$, which is where the dephasing rate reaches its maximum, $\Gamma_{\rm P} \approx 2m^2/\tau_e m_e^2$, while MMM-induced depletion can still be neglected. 
The greatest excluded classicalization time $\tau_{\rm m}$ yielding the macroscopicity value is thus obtained in this regime, and we will restrict our subsequent evaluation to this case.

We now perform the Bayesian hypothesis test with the data taken from the Stanford atom fountain experiment \cite{kovachy2015}, which claims to test the superposition principle on the half-metre scale. The original claim was debated, because it hinged on two crucial assertions  \cite{stamper2016verifying,kovachy2016kovachy}: (i) the rubidium condensate splits coherently and accumulates a stable relative phase $\varphi$ between the two arms in each shot of the experiment, and (ii) uncontrolled vibrations of the recombining beam splitter cause the phase (and the resulting atom counts) to fluctuate randomly from shot to shot. The authors estimated from the data of several dozen shots 
values of an average interference contrast and phase that parametrize the expected atom count distribution, but these estimates alone are not a sufficient criterion for a coherently split condensate state \cite{schrinski2017sensing}. Recent comparative studies have largely ignored this issue \cite{fein2019quantum,carlesso2019collapse}.

To illustrate our formalism and assess macroscopicity on the half-metre scale, we shall assume (i), but \emph{not} (ii). The reason is that, if we knew (ii) were true, we would be prompted to describe the measurement statistics by the phase-averaged distribution \eqref{eq:discreteBECdecohered}, which is insensitive to any further MMM-induced dephasing and thus unamenable to our macroscopicity analysis.  Instead, we take the position of an observer that is uninformed about the phase noise and specify that the data be described by the count distribution \eqref{eq:partdec} subject to MMM dephasing, with a \emph{fixed} unknown phase $\varphi$ that we determine \emph{a posteriori} as the one that maximizes the resulting macroscopicity.

\begin{figure}
  \centering
  \includegraphics[width=0.45\textwidth]{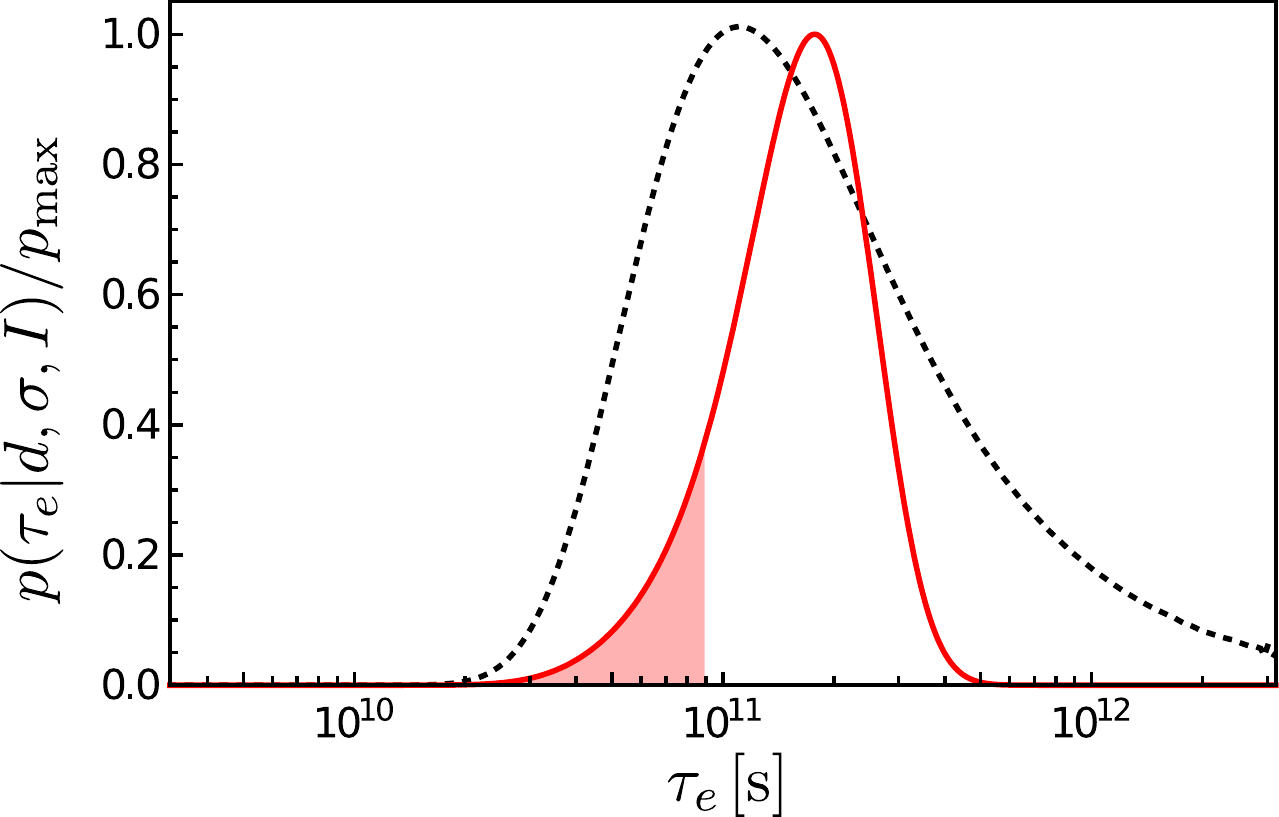}
  \caption{Analysis of the Stanford BEC interferometer \cite{kovachy2016kovachy}: Jeffreys' prior based on the likelihood \eqref{eq:partdec} is shown as the dotted curve and updated with the data at $90\hbar k$ transferred photon recoils. Optimizing the macroscopicity over all possible relative phases $\varphi$ results in the solid curve at $\varphi_\mathrm{max}\approx 6$, which updates the prior to slightly larger $\tau_e$. The shaded region marks the lowest five percent quantile that sets the macroscopicity to $\mu_{\rm m} = 10.9$. Both distributions are normalized to peak value $p_\mathrm{max}$. }
\label{fig:Kasevich}
\end{figure}

Figure \ref{fig:Kasevich} shows the relevant posterior probability for $\tau_e$ (solid line) resulting from 20 data points at $\Delta_p = 90\, \hbar k$ momentum splitting and $t=2.08\,$s, i.e.~at an effective arm separation of $\Delta_x\simeq29\,\mathrm{cm}$. The lowest 5\% quantile yields $\mu_m=10.9$. Compared to the solid lines in Figs.~\ref{fig:KDTLI} and \ref{fig:LUMI}, the Bayesian update neither shifts nor narrows the posterior here significantly with respect to Jeffreys' prior (dotted line). This lack of convergence stems from the lack of reproducible data: the posterior is based on a mere 20 data points, whose phase information is scrambled by the uncontrolled noise we had to neglect. A conclusive test of macrorealistic dephasing on the half-metre scale would require additional measurements. They could either reveal the remaining phase information, in which case the posterior would approach a narrow Gaussian distribution around a $\tau_e$-value that matches the observed phase losses. Or, if the observed phase is indeed random, the posterior would be pushed towards smaller $\tau_e$ than expected a priori.

\subsection{Nested Mach-Zehnder BEC interferometry}

Such additional measurements 
of the phase
are not required in the subsequent nested  Mach-Zehnder experiment reported by the Kasevich group in Ref.~\cite{asenbaum2017phase}. 
They measured phase shifts induced by tidal forces in a nested dual Mach-Zehnder setup, in which they apply a sequence of laser-driven Bragg transitions to split a BEC of $N \simeq 10^6$ Rb atoms at $50\,$nK into two arms separated by $\Delta_p = 102 \hbar k$. They then form identical Mach-Zehnder interferometers as in Fig.~\ref{fig:MachZehnder} in each arm by additional splitting and recombination stages at $\Delta_p = 20\hbar k$. Finally, instead of recombining the output ports of the two outer arms, the authors image the two atom clouds in each single shot and read out their relative phases using a phase-shear technique \cite{sugarbaker2013enhanced}: a sinusoidal density modulation is imprinted onto the two split components, which shows up as spatial fringe patterns in their fluorescence images, with an offset determining the phase difference. This measurement configuration suppresses the effect of vibration-induced fluctuations of the beam splitter phase. 

Ideally, all atoms occupy the same nested two-mode superposition state,
\begin{align}
|\psi\rangle=\frac{1}{\sqrt{2^N N!}}\left(\mathsf{c}_1^\dagger(\varphi_1)+e^{i\varphi_0}\mathsf{c}_2^\dagger(\varphi_2)\right)^N|\mathrm{vac}\rangle,
\end{align}
just before recombination of the inner Mach-Zehnder interferometers. The two arms, spatially separated over 20\,cm at a relative phase $\varphi_0$, are superpositions of two Mach-Zehnder modes with relative phases $\varphi_{1,2}$,
\begin{align}\label{eq:modeDualMZ}
\mathsf{c}_{1,2}^\dagger(\varphi_{1,2})=\frac{1}{\sqrt{2}}\left(\mathsf{a}_{1,2}^\dagger+e^{i\varphi_{1,2}}\mathsf{b}_{1,2}^\dagger\right).    
\end{align}
The creation operators $\mathsf{a}_{1,2}$ and $\mathsf{b}_{1,2}$ respectively create those inner two modes separated by up to 7\,cm. Their relative phases $\varphi_{1,2}$ fluctuate randomly from shot to shot due to uncontrolled vibrations, but their difference $\Delta \varphi$ remains stable. It is extracted in each shot by comparing the phase-sheared images of the two atom clouds. The fluctuating outer phase $\varphi_0$ has no relevance in the following, since the two outer arms are never recombined in the experiment. 

The probability to extract 
a phase value of $\phi_{1}$ from the image in one arm and of $\phi_{2}$ from the other is then simply given by the probability that any $n$ out of $N$ atoms occupy the mode $\mathsf{c}_1 (\phi_{1})$, whereas the other $N-n$ occupy $\mathsf{c}_2 (\phi_{2})$,
\begin{align}\label{eq:NestedMeasurement}
p(\phi_{1},\phi_{2}|\tau,\sigma,I)= \sum_{n=0}^N \langle n,N-n |\rho |n,N-n\rangle,
\end{align}
with 
\begin{align}
|n_1,n_2\rangle=\frac{1}{\sqrt{n_1!n_2!}}\left[\mathsf{c}_{1}^\dagger(\phi_{1})\right]^{n_1}\left[\mathsf{c}_{2}^\dagger(\phi_{2})\right]^{n_2}|\mathrm{vac}\rangle .    
\end{align}
In fact, we can view the $n$ and $N-n$ atoms in the two outer arms as independent condensates.
In the absence of decoherence, we have $\rho = | \psi \rangle \langle \psi |$ and a binomial distribution of the atom portion $n$ that is sharply peaked around $N/2$. We incorporate MMM-induced dephasing by integrating the master equation \eqref{eq:Superoperator} over the interference time $t$ of the two simultaneous Mach-Zehnder stages. Neglecting atom losses and phase dispersion due to atom-atom interactions, we can treat the two branches separately and in the same manner as before. 
To simplify further, we make use of $N\gg 1$ by setting $n\approx N/2$ in \eqref{eq:NestedMeasurement} and performing the continuum approximation. This leaves us with $p(\phi_{1},\phi_{2}|\tau,\sigma,I) \approx p(\phi_{1}|\tau,\sigma,I)p(\phi_{2}|\tau,\sigma,I)$ and an approximately Gaussian phase distribution in each branch that is smeared out by the MMM-induced dephasing around the actual phase value, \begin{align}\label{eq:SingleLikelihood}
&p(\phi_{1,2}\in[-\pi,\pi]|\tau,\sigma,I)\approx\frac{1}{2\pi}\vartheta_3\left(\frac{\phi_{1,2}-\varphi_{1,2}}{2},g_{N/2}(t)\right)\nonumber\\
=&\sum_{k=-\infty}^\infty\mathcal{N} \left( \phi_{1,2}+2\pi k \bigg| \varphi_{1,2}, \frac{1}{N} + \frac{\Gamma_{\rm P} t}{2} \right).
\end{align}
It is the conjugate of the dephased number distribution \eqref{eq:partdec} obtained by re-substitution of $\sin\phi=2n_a/N-1$ and a subsequent $-\pi/2$ pulse (For more details see Ref.~\cite{schrinski2019macroscopicity}). Here, $\mathcal{N}(x|\mu,\Delta)$ stands for a normalized Gaussian distribution with mean $\mu$ and variance $\Delta$ and the sum accounts for the fact that the MMM-induced decoherence may smear the initially narrow distribution beyond the periodic interval $(-\pi,\pi)$.

\begin{figure}
  \centering
  \includegraphics[width=0.45\textwidth]{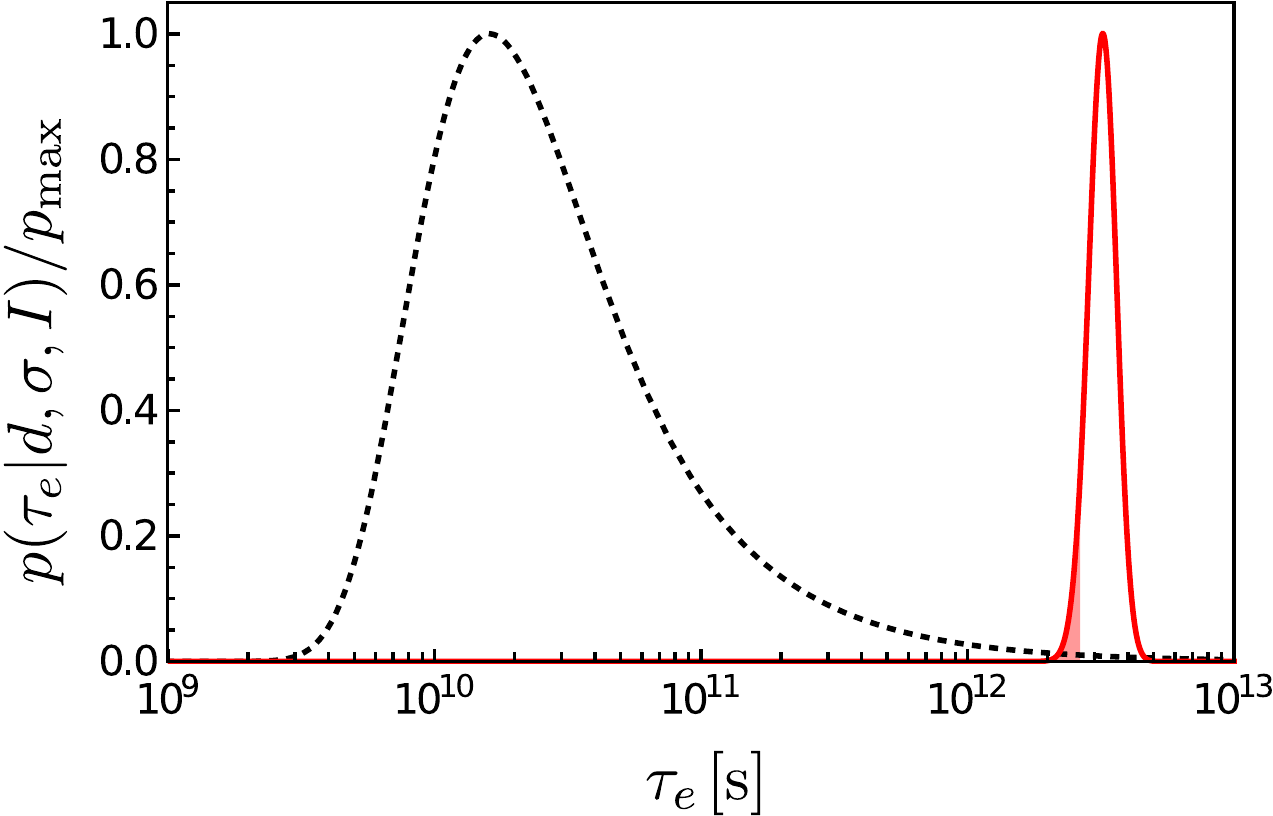}
  \caption{ Analysis of the Stanford nested BEC interferometer \cite{asenbaum2017phase}:
  Jeffreys' prior based on the likelihood \eqref{eq:DualLikelihood} is shown as the dotted curve and updated with the data at $20\hbar k$ transferred photon recoils in both Mach-Zehnder interferometers. The posterior is updated to much greater classicalization times due to the superb localization of the relative phase measured in the experiment. The shaded region marks the lowest five percent quantile that sets  the macroscopicity to $\mu_{\rm m} = 12.4$. Both distributions are normalized to peak value $p_\mathrm{max}$.}
\label{fig:Kasevich2}
\end{figure}

Now we can calculate the probability distribution of $\Delta \phi=\phi_1-\phi_2$, the difference of two random variables, via convolution, which returns once again a Gaussian with double the width and samples the actual interferometer phase difference $\Delta \varphi$ unaffected by beam splitter vibrations,
\begin{align}\label{eq:DualLikelihood}
p(\Delta\phi\in[-\pi,\pi]|&\tau,\sigma,I)=
\frac{1}{2\pi}\vartheta_3\left(\frac{\Delta\phi-\Delta\varphi}{2},g^2_{N/2}(t)\right).
\end{align}
In the experiment \cite{asenbaum2017phase}, the interferometer phase $\Delta\varphi$ was varied by inserting a large test mass in half of the 138 shots recorded at $t=1.2\,$s interference time. However, the actual value of $\Delta\varphi$ is irrelevant, and only the spread of the data points $\Delta \phi$ around the theoretically expected mean matters for our hypothesis test. This spread turns out to be much greater than $1/N$, which implies $N \Gamma_{\rm P} t/2 \gg 1 $ in \eqref{eq:DualLikelihood} and justifies \emph{a posteriori} that we could assume a fixed $N$ and neglect atom number fluctuations in the condensate. 
We arrive at the posterior shown in Fig.~\ref{fig:Kasevich2}, a well converged peak far to the right of the broad prior. Notice the striking improvement over the previous result in Fig.~\ref{fig:Kasevich}, which can be attributed to the fact that the data sample localizes at a stable phase difference. 

\subsection{Interferometry with individual atoms}

In the case of single-atom interference, quantum statistics of identical particles plays no role, 
which greatly simplifies the analysis. We shall demonstrate this by means of a recent experiment with Cs atoms realizing a Mach-Zehnder scheme with fixed arm separation $\Delta_x$ and $t=20$\,s of coherence time \cite{xu2019probing}. Each of the four million recorded atoms is brought into a two-arm superposition that accumulates a variable controlled phase difference $\varphi_k$. At recombination, the state is split into four branches, out of which only two interfere, whereas the other two contribute to the detection signal incoherently and thus halve the interference visibility. MMM dephasing at the rate \eqref{eq:DephRate} would reduce it further  predicting the probabilities
\begin{align}\label{eq:SingleAtomProb}
p_k := p(a|\varphi_k,\tau_e,\sigma,I)=\frac{1}{2}-\frac{\cos(\varphi_k)}{4}e^{-\Gamma_\mathrm{P}t/2},
\end{align}
and $p(b|\varphi_k,\tau_e\sigma,I)=1-p_k$, to detect the atom in the output ports $a$ and $b$, respectively.
The phase difference is varied in 
equidistant steps by adding small submillisecond increments $\delta t \ll t$ to the interference time, $\varphi_k = \omega (t+ k \delta t)$ with $\omega=2\pi\times 12.7\,\mathrm{kHz}$. As $N_k \gtrsim 10^4$ atoms are recorded in each step, we can approximate the resulting binomial distribution for the number $n_a$ of particle counts in $a$ by a Gaussian distribution and take $n_a$ as a continuous variable running from zero to $N_k$,
\begin{align}\label{eq:BinomialToGauss}
p(n_a|\varphi_k,N_k,\tau_e,\sigma,I) &\simeq \mathcal{N}(n_a|N_k p_k,N_k p_k(1-p_k))\nonumber\\
&\propto\exp\left[-\frac{(n_a-N_k p_k)^2}{2 N_k p_k(1-p_k)}\right].
\end{align}
This approximation for the likelihoods of detecting $n_a$ atoms in $a$ would be inaccurate close to the boundaries $n_a=0,N_k$, but these extreme values  never occur in practice  in a realistic low-contrast scenario.

From Eq.~\eqref{eq:SingleAtomProb} we infer that, whenever $\varphi_k$ is an odd multiple of $\pi/2$, the likelihoods $p(n_a|\varphi_k,N_k,\tau_e,\sigma,I)$ do not depend on the MMM parameters $\tau_e,\sigma$, but instead match the classical coin-flip model of the Mach-Zehnder setup (i.e.~a binomial count distribution centered at $n_a=N/2$). Data points recorded at such $\varphi_k$ are thus useless in terms of macroscopicity, as they do not update the posterior of the MMM time parameter $\tau_e$. 
Notice that the count distribution matches the classical model also in the limit of complete dephasing (unlike the BEC interferometer case, see Eq.~\eqref{eq:discreteBECdecohered}).

\begin{figure}
  \centering
  \includegraphics[width=0.45\textwidth]{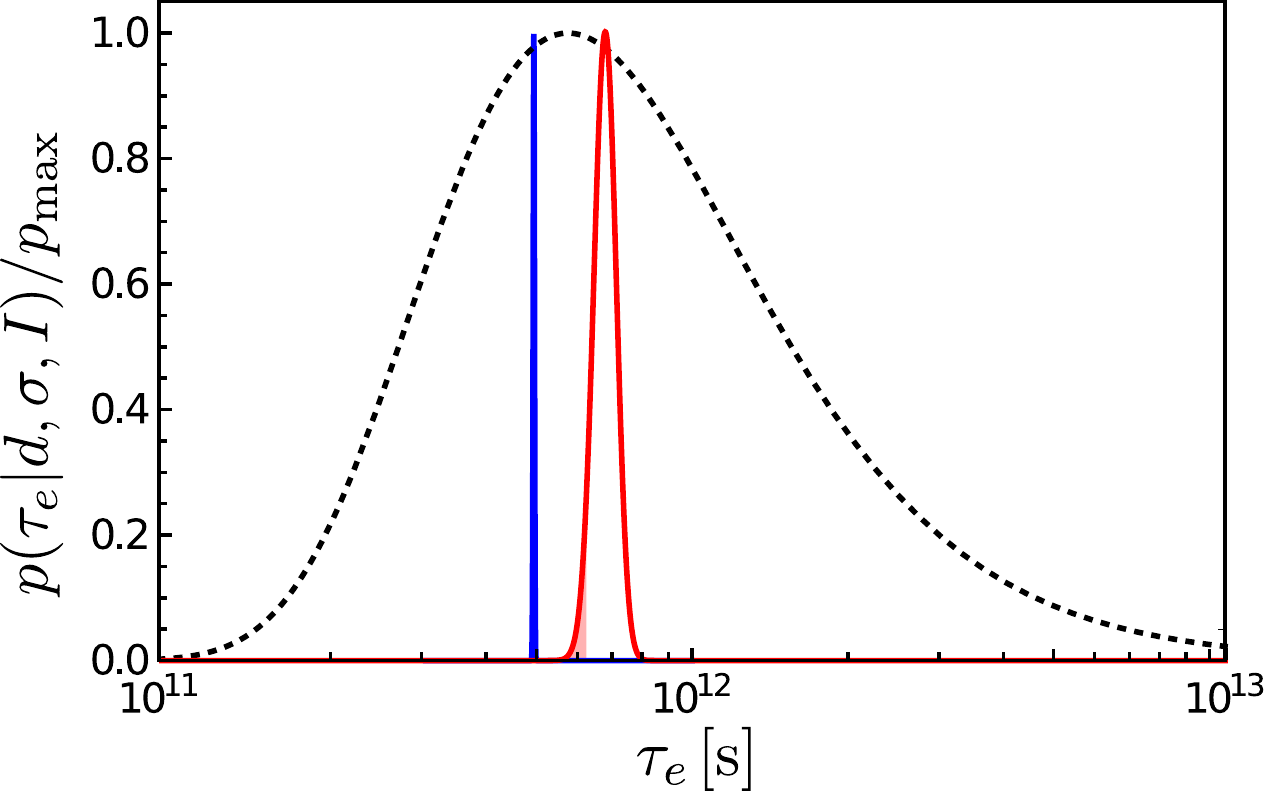}
  \caption{Analysis of the Berkeley atom interferometer \cite{xu2019probing}: Jeffreys' prior based on the likelihood \eqref{eq:BinomialToGauss} (dotted line) is updated with the atom count data at $t=20\,\mathrm{s}$ interference time, resulting in a quasi-$\delta$ posterior (blue solid line). A more realistic assessment including fluctuating dark counts with  $\sigma_\mathrm{dark}=10^3$ standard deviation does not change the prior appreciably, but yields a broader posterior (red solid line) with a higher macroscopicity
  $\mu_{\rm m}=11.8$, as marked by the shaded region. Once again, all distributions are normalized to their respective maxima $p_\mathrm{max}$.}
\label{fig:20sec}
\end{figure}

Contrary to some of the previous examples, there is no shortage of data here to do Bayesian inference. Taking all 4 million recorded data points for $t=20\,$s into account, we obtain a sharply peaked posterior for $\tau_e$, with a corresponding macroscopicity value $\mu_{\rm m}=11.7$ and a tiny FWHM of $3.5\times 10^{9}\,\mathrm{s}$ (i.e.~less than one percent of the corresponding $\tau_\mathrm{m}$). 
However, such a small error in the hypothetical $\tau_e$-estimate seems ``too good to be true'', as it implies perfect single-atom detection for all four million registered counts,
which is unrealistic.

For this specific experiment, the most prevalent noise source are dark counts in the CCD cameras; these instruments are supposed to detect atoms via fluorescence imaging, but occasionally they miss an atom or click when there is none. In conventional data analysis, one would subtract an appropriate background level from the average detection signal and top up its error bar. In Bayesian inference, we must introduce a random variable and convolute our model likelihoods with the respective noise distribution. The most obvious choice, a Gaussian random variable, is easily incorporated in our Gaussian approximation \eqref{eq:BinomialToGauss} of the likelihoods by adding its contribution $\sigma^2_\mathrm{dark}$ to the overall variance, i.e.~writing $\mathcal{N}(n_a|N_k p_k,N_k p_k(1-p_k)+\sigma^2_\mathrm{dark})$. The corresponding Jeffreys' prior is given in App.~\ref{AppA}. For the present experiment, a realistic lower bound for the dark count fluctuations would be $\sigma_{\rm dark} \simeq 10^3$. The resulting posterior yields the macroscopicity $\mu_{\rm m} = 11.8$ at a reasonable FWHM uncertainty of $7.85\cdot10^{10}\,\mathrm{s}$. 

Figure \ref{fig:20sec} compares the posterior (red solid line) to the one omitting dark counts (blue line, almost $\delta$-peaked) and to Jeffreys' prior (dotted line). Both posteriors have not moved far from the prior, which is consistent with the low interference contrast reported in \cite{xu2019probing}.  
We observe that the inclusion of dark counts not only leads to a more realistic spread (i.e.~uncertainty) of the distribution, but also to a systematic shift towards greater $\tau_e$ and $\mu_{\rm m}.$  This exemplifies how measurement errors and their proper statistical assessment affect the empirical macroscopicity of a quantum experiment and its capability to rule out macrorealistic modifications such as CSL. The decohering effect of any technical or environmental noise source that one knowingly omits would be attributed to MMM, which overestimates their hypothetical strength, i.e.~underestimates $\tau_e$ and $\mu_{\rm m}$.

\section{Convergence of the posterior distribution}\label{sec:Hellinger}

The Bayesian approach to hypothesis testing 
utilized in this paper not only 
permits comparing matter-wave experiments in terms of their macroscopicity; it  also reveals, through the observed degree of posterior convergence, how conclusive the measurement records are at testing macrorealism. This can be turned into a quantitative statement by employing Bayesian consistency and 
tools from probability theory.

Suppose we interpret 
the  Bayesian updating as a parameter estimation method for a specific macrorealistic model such as CSL, with the aim of finding the ``true'' underlying time parameter $\tau_e = \tau_0$ at fixed $\sigma$. For this case of a one-dimensional parameter space and a strictly positive prior it is known
to which  distribution the posterior will converge for an asymptotically large data set
\cite{schwartz1965bayes,vaart1996weak,le2012asymptotics}.
It is the Gaussian $\mathcal{N}(\tau_e|\tau_0,1/\mathcal{F}(\tau_0|\sigma,I))$,  centered around the true parameter value $\tau_0$, whose variance is given by the inverse of the Fisher information  appearing in Jeffreys' prior \eqref{eq:JeffreysPrior}. This suggests  
to assess the degree to which a quantum experiment is a conclusive test of MMM by quantifying how well the respective posterior has converged to that asymptotic distribution. 
It is then natural to measure the residual deviation by means of the (bounded and symmetric)
Hellinger distance \cite{ali1966general}, 
\begin{align}\label{eq:HD}
H^2(d,\sigma,I)=&\frac{1}{2}\int_0^\infty d\tau_e\,\left[\sqrt{\mathcal{N}(\tau_e|\tau_0,1/\mathcal{F}(\tau_0|\sigma,I))}\right.\nonumber\\
&\left.-\sqrt{p(\tau_e|d,\sigma,I)}\right]^2 \leq 1.
\end{align} 

Of course, a true value of $\tau_e$ is not available in practice
because the existence of MMM cannot be verified experimentally. The posterior can only serve to \emph{falsify} collapse models with values of $\tau_e$
up to a threshold
(e.g. based on the lowest 5\% quantile,
as in the empirical definition of macroscopicity). 
After all, an observed reduction of interference visibility 
might always be caused by uncontrolled technical disturbances, or due to an unidentified environmental source of decoherence. 
Given that the posterior expectation value of $\tau_{\rm e}$ may not exist, we take the minimum of \eqref{eq:HD} over all possible $\tau_0$ as a pragmatic measure for posterior convergence. Small values thus indicate more conclusive experimental tests of MMM at the given macroscopicity level.

\begin{table}
\begin{tabular}{|| c| c |c |c | c||}
\hline
Experiment & FWHM & Gaus.~FWHM & min.~HD & $\mu_{\rm m}$ \\ \hline
KDTLI \cite{eibenberger2013}& $1.80\cdot10^{11}\,\mathrm{s}$ & $1.77\cdot10^{11}\,\mathrm{s}$ & $0.00054$ & 12.3 \\  
LUMI$_8$ \cite{fein2019quantum}& $1.94\cdot10^{15}\,\mathrm{s}$ & $6.66\cdot10^{15}\,\mathrm{s}$ & 0.52 & 14.8\\  
LUMI$_{23}$ \cite{fein2019quantum}& $2.63\cdot10^{13}\,\mathrm{s}$ & $2.76\cdot10^{13}\,\mathrm{s}$ & 0.079 & 14.0\\ 
MZI(BEC) \cite{kovachy2015}& $1.69\cdot10^{11}\,\mathrm{s}$ & $1.74\cdot10^{11}\,\mathrm{s}$ & 0.13 & 10.9\\  
nMZI(BEC) \cite{asenbaum2017phase}& $9.22\cdot10^{11}\,\mathrm{s}$ & $9.28\cdot10^{11}\,\mathrm{s}$ & 0.0012 & 12.4\\  
MZI(atoms) \cite{xu2019probing}& $7.85\cdot10^{10}\,\mathrm{s}$ & $7.69\cdot10^{10}\,\mathrm{s}$ & 0.018 & 11.8\\
\hline
\end{tabular}
\caption{Survey of posterior convergence and macroscopicity for the evaluated experimental scenarios: We compare the full widths at half maximum (FWHM) of the posteriors with those of the respective asymptotic Gaussians that minimize the Hellinger distance \eqref{eq:HD} (min.~HD). The right column lists the corresponding macroscopicities. \label{tab:HD} }
\end{table}

Table \ref{tab:HD} lists the minimal Hellinger distances for the discussed 
experiments, together with the FWHM widths of the posteriors and of the corresponding asymptotic Gaussians. The KDTLI data, the LUMI data including all 23 interferograms, the Stanford nested interferometer, and the Berkeley Mach-Zehnder interferometer result in Hellinger distances much less than unity, indicating a conclusive test of macrorealism. On the other hand, the best eight interferograms of LUMI and the 20 data points from the half-meter BEC interferometer at Stanford yield values not far from unity, which indicates poor posterior convergence and suggests that more data be taken for a conclusive result. This is further indicated by the Gaussian FWHM being of the same order as the mean in both cases, i.e.~the asymptotic distribution would extend appreciably to unphysical values $\tau_e<0$.  Taking these considerations into account the LUMI experiment should be associated with a macroscopicity  of $\mu_\mathrm{m}=14.0$, rather than the $\mu_\mathrm{m}=14.8$ derived from the reduced data set. 

\section{Conclusion}\label{sec:Conclusion}

\begin{figure}
  \centering
  \includegraphics[width=0.45\textwidth]{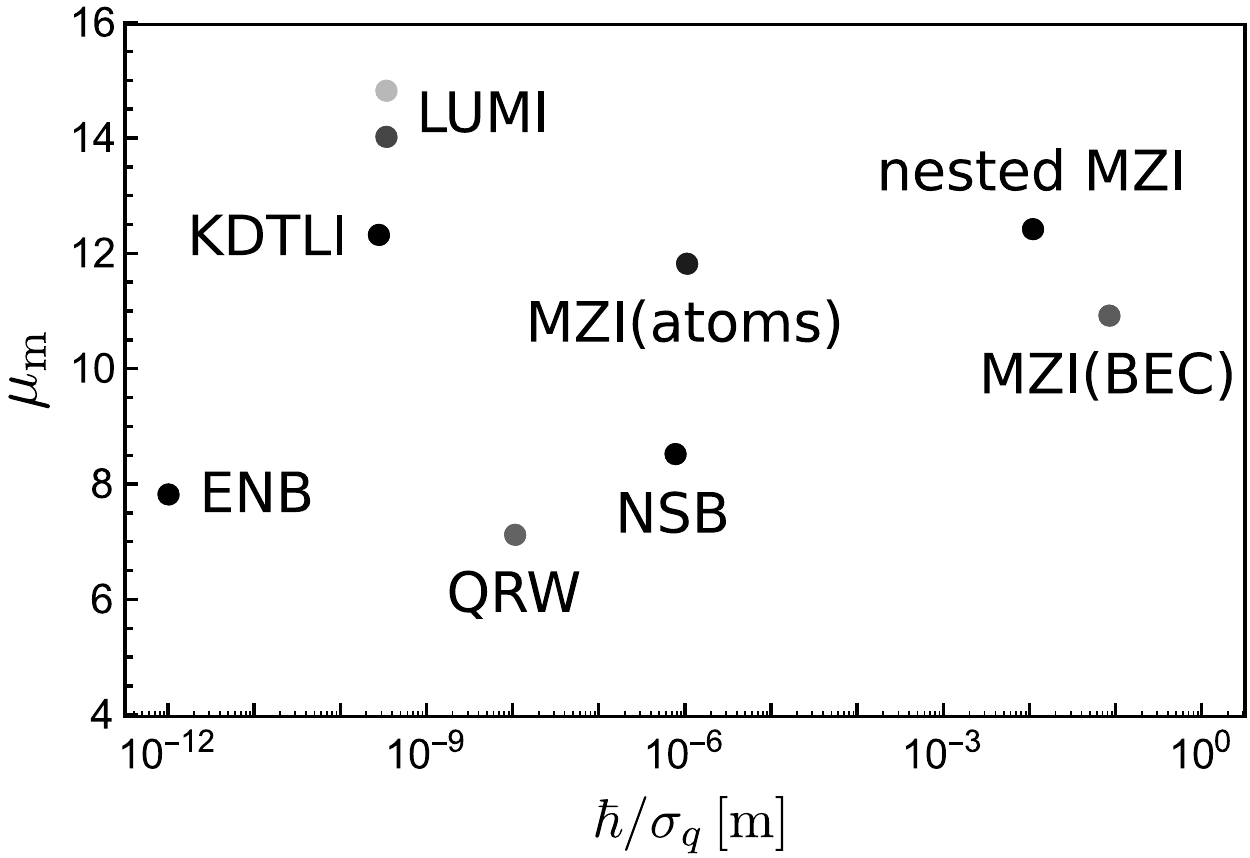}
  \caption{Macroscopicities of the experiments evaluated in this paper and in Ref.~\cite{schrinski2019macroscopicity}. This includes entangled nanobeams (ENB \cite{riedinger2018}), quantum random walks (QRW \cite{Robens2015}), number-squeezed BEC interferometry (NSB \cite{Schmiedmayer2013}), molecule interferometry (KDTLI \cite{eibenberger2013} and LUMI \cite{fein2019quantum}), Mach-Zehnder interferometry (MZI) with single atoms \cite{xu2019probing}, with BECs \cite{kovachy2015}, and nested BECs \cite{asenbaum2017phase}. 
  The values are decadic logarithms \eqref{eq:mu} of the greatest falsified time parameters of minimal macrorealistic models, plotted against the greatest length scale at which the experiments probe macrorealism effectively \cite{footFig}. The brightness of the points is directly proportional to the respective Hellinger distance, as presented in Sec.~\ref{sec:Hellinger}. }
\label{fig:TauVsSigma}
\end{figure}

In this paper, we analyzed 
the most recent matter-wave interference experiments with atoms, molecules, and BECs regarding their capability to probe the quantum-classical transition by ruling out minimal macrorealistic modifications (MMM) of quantum theory. 
Building on a parametrization of this class of 
models, Bayesian parameter inference with Jeffreys' prior allows one to  assess and objectively compare the different  experiments. The achieved macroscopicity $\mu_{\rm m}$ is then given by the
maximal classicalization time parameter
ruled out by the measurement data  at a significance level of 
5\%.
Based on this yardstick, the matter wave interference reported in Refs.~\cite{eibenberger2013,asenbaum2017phase,fein2019quantum,xu2019probing} yield the highest macroscopicities demonstrated in any test of the quantum superposition principle so far. 

A second quantity characterizing the 
pertinence of individual interferometers is the greatest critical length $\hbar/\sigma_q$ for which the falsified classicalization time is maximized. At this scale, often on the order of the interference path separation, the instrument is  most sensitive to collapse effects on delocalized quantum states.
Figure \ref{fig:TauVsSigma} shows the macroscopicity against this 
critical length, comparing the discussed interference experiments with previous case studies. 
One observes that the scales probed by different superposition tests vary by ten orders of magnitude. We remark that the BEC interference experiment by the Kasevich group reaches
$\hbar/\sigma_q=8$\,cm, about an order of magnitude less than the geometric path separation of up to half a meter; this is because the effective splitting distance $\Delta_x=29$\,cm has to be sufficiently undershot to maximize the classicalization effect, as is the case for all interference experiments shown here. 

We note that the Bayesian method demonstrated in this paper can be readily applied to constrain all the parameters of a specific collapse model such as continuous spontaneous localization (CSL). By combining data from distinct experiments one would obtain a probability distribution on the set of all model parameters that can be successively sharpened  by Bayesian updating. 
However, an uninformative prior that is invariant under reparametrization is not available for a multi-parameter space \cite{bernardo1979reference} so that in this case one must rely on the posterior turning independent of the prior once sufficient data has been processed.  An exemplary result with the experiments discussed in this paper is shown in App.~\ref{AppB}. A great advantage of this Bayesian parameter estimation, which builds on the raw data of each experiment, is that the measurement statistics is correctly accounted for. This is in contrast to naive exclusion plots based on average values for each experiment in a frequentist fashion, for which it is very hard to carry out a consistent error analysis.

Apart from demonstrating the significance of statistical errors in diverse types of measurement data,
our analysis also highlighted the importance of a realistic modeling of the experiment including technical noise and environmental decoherence, and the pitfalls of post-selecting a subset of ``successful runs''.
As the size and complexity of experiments exploring the quantum-classical boundary will grow, so will the difficulty to keep errors under control and collect a sufficient amount of reliable data. The presented Bayesian protocol provides a rigorous method to assess the quality of the data in regard to probing macrorealism and to objectively quantify the degree of macroscopicity. It could be readily applied to space-borne precision interferometry \cite{altschul2015quantum,kaltenbaek2016macroscopic,becker2018space} that is 
projected to outperform the current record experiments in the future.

\acknowledgments
We would like to thank S.~Eibenberger, Y.~Fein, and V.~Xu for providing us with their  experimental data. This work was funded by Deutsche Forschungsgemeinschaft (DFG, German Research Foundation), Grant No. 298796255.

\appendix
\begin{widetext}
\section{Jeffreys' prior for single-atom interferometry}\label{AppA}  

The experimental setting of single-atom Mach-Zehnder interferometry subject to MMM is one of the rare instances where Jeffreys' prior \eqref{eq:JeffreysPrior} can be expressed in compact form, in the limit when the binomial distribution of atom counts is well approximated by a Gaussian with a continuous number variable $n_a$, cf.~\eqref{eq:BinomialToGauss}. We consider the scenario where the interference phase is varied over discrete values $\varphi_k$ and $N_k\gg1$ atoms are counted for each phase value. The approximation is then safely valid for the present count numbers beyond several thousands as long as the interference contrast does not reach values close to 100\%. Given that the Gaussian distribution for $n_a \in [0,N_k]$ then practically vanishes at the interval boundaries, we may extend the integration to $n_a \in \mathbb{R}$ and write

\small
\begin{align}\label{eq:AtomJeffreysPrior}
p(\tau_e|\sigma,I) \propto \sqrt{\sum_k\int_{-\infty}^\infty dn_a\,\frac{[\partial_{\tau_e}\mathcal{N}(n_a|N_k p_k,N_k p_k (1-p_k)+\sigma^2_\mathrm{dark})]^2}{\mathcal{N}(n_a|N_k p_k,N_k p_k (1-p_k)+\sigma^2_\mathrm{dark})}}
\simeq\sqrt{\sum_k\frac{2N_k^2\tilde{\Gamma}_\mathrm{P}^2t^2\cos^2(\varphi_k) e^{\tilde{\Gamma}_\mathrm{P}t/\tau_e}}{\pi\tau_e^4[N_k\cos^2(\varphi_k)-4(N_k+4\sigma_\mathrm{dark}^2)e^{\tilde{\Gamma}_\mathrm{P}t/\tau_e}]^2}},
\end{align} 
\normalsize
with $\Gamma_\mathrm{P}=\tilde{\Gamma}_\mathrm{P}/\tau_e$. Here we have omitted the normalization constant, but have incorporated the possibility of Gaussian-distributed dark counts adding, the contribution $\sigma_{\rm dark}^2$ to the overall variance of the individual count distributions.

 The square of \eqref{eq:AtomJeffreysPrior} is the Fisher information about the $\tau_e$-parameter contained in the Gaussian likelihood. In Fig.~\ref{fig:BECvsSingleAtoms}(a)  we compare it to the Fisher information associated with a corresponding interference experiment with Bose-Einstein-condensed atoms.  In both cases the atom number is chosen as $N=10^3\gg 1$, so that the continuum approximation applies (dark counts noise is omitted, $\sigma_{\rm dark}=0$). 
One observes that the BEC setup is slightly more sensitive to weaker modifications (i.e.\ higher classicalization time parameters $\tau_{\rm e}$), illustrating the prospect to gain higher macroscopicities.  Note however that measurement inaccuracies and decoherence effects of real experimental setups are not accounted for. Panel (b) shows the initial and the fully classicalized atom number distributions in one output mode. In case of individual atoms it approaches the binomial distribution expected for independent coin flips, while it approaches  the intensity distribution of a random classical wave in the case of a BEC.

\begin{figure}
  \centering
  \includegraphics[width=0.99\textwidth]{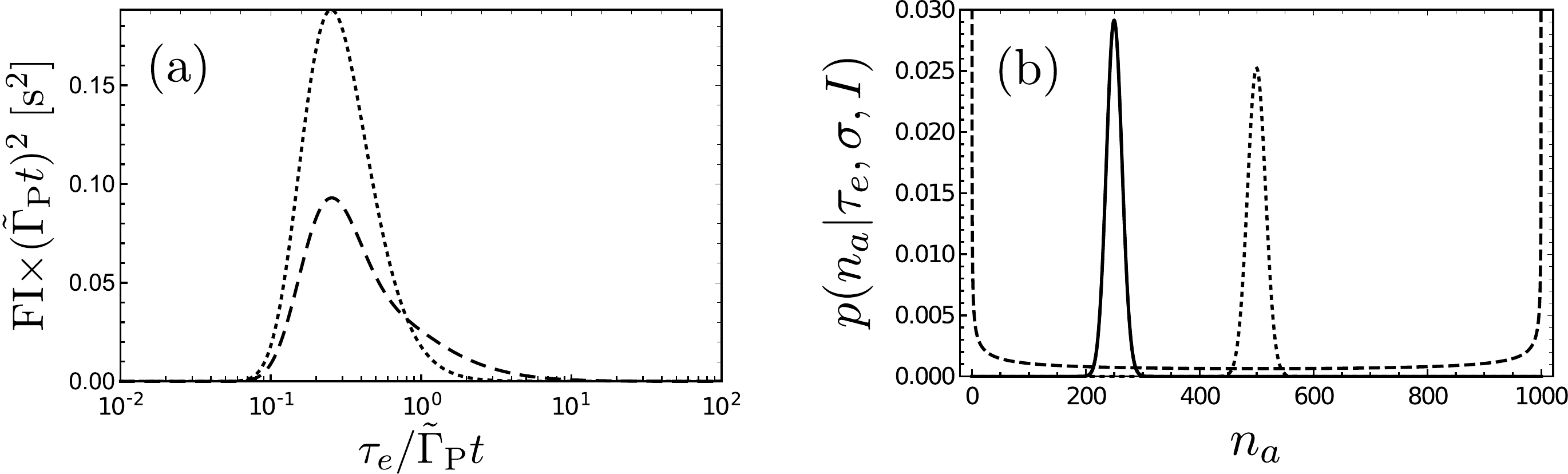}
  \caption{ (a) Fisher information of the interference experiment  with individual atoms (dotted line), as determined by the square of the unnormalized Jeffreys' prior \eqref{eq:AtomJeffreysPrior}, and with Bose-Einstein-condensed atoms (dashed line), with $N=1000$ atoms each. (b) Corresponding atom number distributions in one of the output modes. 	In absence of a modified dynamics the distributions coincide (solid line), while they assume the behavior expected for random classical particles (individual atoms, dotted) and waves (BEC, dashed), respectively, in the fully decohered case. }
\label{fig:BECvsSingleAtoms}
\end{figure}

\section{Bayesian updating from experiment to experiment}\label{AppB}

One strength of Bayesian inference is the effortless combination of different experiments to an overall picture by simply multiplying the likelihoods of the realized experimental observations. That is, one can use the posterior of one experiment as prior for another one.
In Fig.~\ref{fig:GesamtPost} (top row, (a)-(e)), we plot the likelihoods $p(d|\tau_e,\sigma_q,I)$ resulting from each experiment discussed in this paper in the two-dimensional parameter space $(\tau_e,\sigma_q)$. Each distribution exhibits a flat plateau where the decoherence rate saturates for small $\hbar/\sigma_q$, yielding the maximum $\tau_e$ that determines the respective  macroscopicity. The transition to the diffusive regime at large $\hbar/\sigma_q$, where all distributions exhibit a quadratic scaling with $\sigma_q$, typically takes place when $\hbar/\sigma_q$ matches roughly the path separation of the involved superpositions. 

Here we illustrate how the results of the discussed experiments  combine into a joint parameter estimate of macrorealism, regardless of whether the observations are genuine quantum interference or particle loss. Assuming that particle loss was monitored sufficiently well, we can neglect that the decoherence rate drops in  the limit $\hbar/\sigma_q\to 0$ (see Sect.~\ref{sec:tmbec}) and extend the plateaus in Fig.~ \ref{fig:GesamtPost}(a)-(e) to $\hbar/\sigma_q\to 0$. The lower panels (f)-(i) illustrate the successively updated combined posterior as subsequent experiments are factored in. The localized distribution in the final panel (i) thus represents the most likely range of MMM parameters based on the five experiments. 

We emphasize that such a parameter estimate is based on the strong assumption that MMM-induced classicalization does exist and that any unidentified source of decoherence or particle loss is attributed to MMM. The combined data thus not only excludes stronger, but also weaker modifications, since systematic errors in the noise assessment are not reflected in this analysis. A reliable procedure to account for those systematic errors would be needed to mimic the exclusion curves prevalent in the literature \cite{carlesso2019collapse}.

Finally, we note that the distributions displayed in Fig.~ \ref{fig:GesamtPost} are obtained by simply multiplying  the likelihoods $p(d|\tau_e,\sigma_q,I)$ of each experiment. Regarding the final distribution (i) as a posterior is equivalent to assuming a flat prior in the very beginning. However, while the likelihoods $p(d|\tau_e,\sigma_q,I)$ for specific $\sigma_q$ are close to the posteriors $p(\tau_e|d,\sigma_q,I)$ in most experiments (as indicated by the Hellinger distances in Tab.\ \ref{tab:HD}), the choice of prior still matters for ensuring integrability and thus normalization. One is then faced with having to choose a more suitable prior, keeping in mind that there exists none that is simultaneously uninformative and independent of re-parametrization in the multi-dimensional parameter space \cite{bernardo1979reference}.

\begin{figure}
  \centering
  \includegraphics[width=0.99\textwidth]{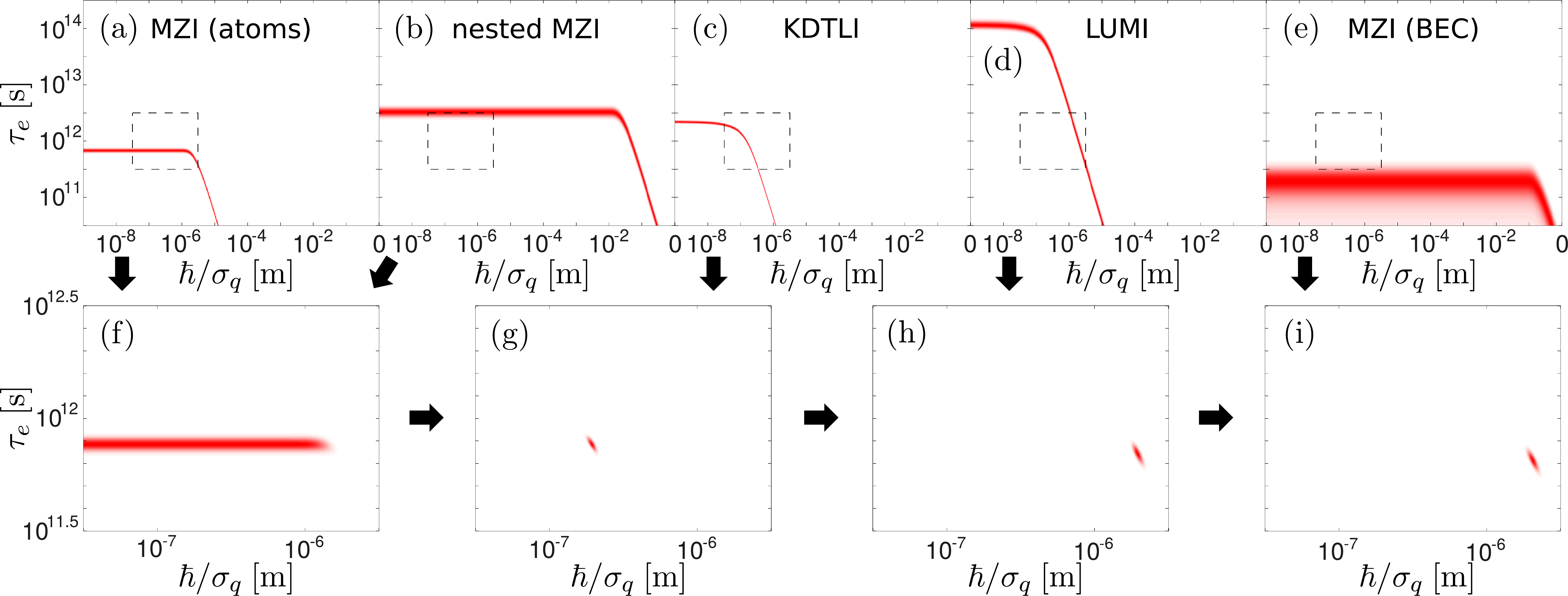}
  \caption{The top row shows the  two-dimensional likelihoods $p(d|\tau_e,\sigma_q,I)$ for the experiments discussed in this paper. The bottom row depicts the combined distributions obtained by accounting for more and more experiments, as indicated by the arrows: (f) If two likelihoods share no significant overlap the combined distribution exhibits an elongated area of maximum likelihood that is cut at $\hbar/\sigma_q\simeq 10^{-6}$--the point where both likelihoods depart. The likelihood of the MZI with atoms dominates the product since it is more peaked. (g) If two likelihoods have overlapping regions the combined distribution peaks at the intersecting region. Here, the different geometries in the involved experiments would pinpoint the most likely modification parameters under the premise that all decoherence effects are indeed a result of MMM. (h) As the likelihood for the LUMI data does not cross the region of the previously combined distribution the latter is updated strongly to a different region. (i) Such an update is negligible if the experiment provides only few data points as in the case of the MZI with BEC. Note that the top and the bottom panels depict different parameter regions (indicated by the dashed rectangles). When comparing the plots to CSL exclusion curves one has to set $\lambda_{\rm CSL}=m_n^2/m_e^2\tau_e$ with $m_n$ the mass of a neutron and $r_{\rm CSL}=\hbar/\sqrt{2}\sigma_q$.}
\label{fig:GesamtPost}
\end{figure}

\end{widetext}


%

\end{document}